\documentclass[10pt, fleqn]{article}

\usepackage[paper=letterpaper, margin=1.5in]{geometry}

\usepackage{setspace} 
\usepackage[T1]{fontenc}
\usepackage{lmodern} 
\usepackage{sourcecodepro}
\usepackage[osf]{mathpazo}
\newcommand{\strong}[1]{\textsf{\textbf{#1}}}

\usepackage{hyperref}
\usepackage{listings}

\usepackage{amsmath}
\usepackage{amssymb}
\usepackage{amsthm}
\newtheorem{example}{Example}
\newtheorem{remark}{Remark}
\usepackage{mathrsfs}
\usepackage{verbatim}
\usepackage{algorithm}
\usepackage{algorithmic}
\usepackage{wasysym} 
\usepackage[pdftex]{graphicx}
\usepackage[pdftex,usenames,dvipsnames]{color}
\usepackage{colortbl} 
\usepackage[table]{xcolor} 
\usepackage{ifmtarg}
\usepackage{boxedminipage}
\usepackage[numbers, sort]{natbib} 
\usepackage{pfResV10}

\newcommand{\parspc}{\vspace{0.6\baselineskip}}

\lstdefinelanguage{PES}
{morekeywords={\#define,define,clocks,control,predicate,start,initially,equations,invariant,transitions},
sensitive=false,
morecomment=[l]{//},
morecomment=[s]{/*}{*/},
morestring=[b]",
}

\title{\textbf{Timed Automata Benchmark Description}\thanks{Research supported by NSF grants CCF-0820072 and CCF-0926194.}}
\author{Peter Fontana and Rance Cleaveland}
\date{\today}

\graphicspath{{Figures/}}

\begin{document}

\newlength{\linenumwidth} \setlength{\linenumwidth}{20pt}
\makeatletter
\def\lst@PlaceNumber{%
  \makebox[\linenumwidth][l]{\normalfont\lst@numberstyle{\thelstnumber}}%
}
\makeatother

\maketitle
\thispagestyle{empty}
\begin{abstract}
This report contains the descriptions of the timed automata (models) and the properties (specifications) that are used as the ``benchmark examples in Data structure choices for on-the-fly model checking of real-time systems'' and ``The power of proofs: New algorithms for timed automata model checking.'' The four models from those sources are: CSMA, FISCHER, LEADER, and GRC. Additionally we include in this report two additional models: FDDI and PATHOS. These six models are often used to benchmark timed automata model checker speed throughout timed automata model checking papers.
\end{abstract}

\tableofcontents

\newpage

\section{Overview of Protocols}

This report describes some of the benchmarks used in \cite{Fontana:2014aa, fontana-data-structure-2011}. to assess the performance of different timed-automaton model-checking algorithms.  Our intention is to document both the models and formulas used in some detail so that other researchers can use them for their own purposes.

	In general, the examples are used in \cite{zhang-model-checking-2005}, which were based on the examples used in \cite{wang-efficient-verification-2003} and the RED (\url{http://cc.ee.ntu.edu.tw/~farn/red/red.5.0.linux.tar.gz}) and updated RED code (\url{http://sourceforge.net/projects/redlib/} .  However, various sources were consulted (sources cited for each benchmark).  When modeling differences existed, we mostly followed \cite{wang-efficient-verification-2003} but we sometimes changed the event names (though not the transitions allowed) to adapt the models to meet the baseline notion of timed automata in \citet{fontana-a-menagerie-2014}.

	The following benchmarks are described:
	
\begin{description}
	\item[CSMA/CD] \emph{Carrier Sense, Multiple Access with Collision Detection}.  This involves $n$ processes who wish to send a message on $1$ bus, where the bus can only handle one message at a time.
	\item[FISCHER] \emph{Fischer's Mutual Exclusion}.  This involves $n$ processes that wish to access a critical section.  They access a critical section by first assigning their id to a global variable, which serves as a lock.
	\item[GRC] \emph{Generalized Railroad Crossing}. This describes $n$ trains (``processes'') who wish to cross train tracks that is gated. Each process is on a different train track, but shares a gate for all tracks.  There is a gate and a controller that synchronize the raising and lowering of the gate with the crossing of the trains.
	\item[LEADER] \emph{Leader Election Protocol}.  This involves $n$ processes that wish to elect a leader among themselves.  They choose a leader by requesting fellow processes (in an asynchronous manner) to be their parent until only on process has no parent.

\end{description}

For each model we model-checked one valid safety specification ($as$), one invalid safety specification ($bs$), one valid liveness specification ($al$), and one invalid liveness specification ($bl$). Each of these cases involves only one temporal operator: $\phi_1$ involves conjunctions and disjunctions of atomic propositions and clock constraints. In addition we tested 4 additional specifications on each property ($M1$, $M2$, $M3$, and $M4$). Out of these specifications, at least one (usually $M4$) is a property with no known equivalent TCTL formula. The specifications checked are listed below. The specifications that are not supported by UPPAAL are in \emph{italics} and are marked with a $^{*}$.

	For the specifications, they will be documented in the following way:
\begin{itemize}
	\item \textbf{AS Example: Valid Safety Specification}: This describes a safety specification that is expressed via a safety property that the system does satisfy.  These specs are the A examples taken from \cite{zhang-model-checking-2005}.
	\item \textbf{BS Example: Invalid Safety Specification} This describes a safety specification that the system does not satisfy. These specs are the B examples taken from \cite{zhang-model-checking-2005}.
	\item \textbf{AL Example: Valid Liveness Specification}: This describes a liveness specification that is expressed via a safety property that the system does satisfy. 
	\item \textbf{BL Example: Invalid Liveness Specification} This describes a liveness specification that the system does not satisfy. These specs are the B examples taken from \cite{zhang-model-checking-2005}.
	\item \textbf{M1 Example:} This provides an additional formula that the model may or may not satisfy. This formula may or may not be expressible in TCTL.
	\item \textbf{M2 Example:} This provides an additional formula that the model may or may not satisfy. This formula may or may not be expressible in TCTL.
	\item \textbf{M3 Example:} This provides an additional formula that the model may or may not satisfy. This formula may or may not be expressible in TCTL.
	\item \textbf{M4 Example:} This provides an additional formula that the model may or may not satisfy. This formula may or may not be expressible in TCTL.
\end{itemize}

	The \textbf{A Example}, \textbf{B Example} and \textbf{C Example} specifications are the ones used in our benchmark suite when testing the model checkers on safety properties.  Unless otherwise specified, when modeling the systems the default parameter values are used for all parameters.
	
	\begin{remark} When either constructing an implementation safety specification, when the system is symmetric for all processes the representation of the specification can be written to only inquire about a few processes, since if it is true for those then it must be true for all processes.  All specifications for the models are written in full because the additional formula results in a changed performance.
	\end{remark}
	
	When specifying the specifications, the TCTL and the $L^{rel}_{\nu,\mu}$ formulas are specified when possible using variables $i,j,k$ to simplify the notation of process conjunctions. For the Tool formulas, the formulas are specified in full but using the fewest number of processes that can be generalized. The number of processes used is two processes unless specified otherwise.

\section{PES Modeling framework}

Here we discuss the structure of a PES (Predicate Equation Systems) model. A Predicate Equation system is a way of combining both a model and a formula together into a system of equations, where the set of states satisfying the equation are the set of states in the model that satisfy the formula. In this report, we encode both a timed automaton and a timed-automaton formula (as a modal equation system) into a PES. 

Modal Equation Systems (MES) express a modal-logic formula as an equation. Timed MES (TMES), exist. One version of Timed MES are in: \cite{sokolsky-local-model-1995}. The version of Timed MES we used are described in \citet{Fontana:2014aa, fontana_towards_2014}.

Here are the different components of a PES Model:

\begin{itemize}
	\item \strong{Comments (//).} Any line starting with ``//'' are ignored by the parser and can be used to provide descriptive text.
	\item \strong{Constants (\# define).} Any line starting with ``\# define'' indicates a name of a constant that can be used throughout the example.
	\item \strong{Clocks ($x_i, y_j$).} The variables after the word ``CLOCKS:'' are the variable names used to indicate all of the timed automata clocks in the model. Clock names begin with an ``x'' or a ``y'' (lowercase) and are (optionally but usually) followed with a number. Unless specified otherwise, all clocks are initialized to $0$.
	\item \strong{State Variables or Atomic Propositions ($p_i, q_j$).} The variables after ``CONTROL:'' are the variable names used for all of the atomic propositions or the state variables. The proposition variables take on non-negative integer variables, and by default are initialized to $0$. Atomic propositions start with a ``p'' or a ``q'' (lowercase) and are (optionally but usually) followed by a number. Optionally, in parentheses, one can specify the number of intended values for each proposition, such as $p_1(2)$. The parentheses are ignored by the parser.
	\item \strong{Predicate Variables ($X_i$).} The variables after the word ``PREDICATE:'' are the predicate or equation variables used in the mu-calculus formula. Each predicate variable starts with an ``X'' (uppercase) and is (optionally but usually) followed with a number.
	\item \strong{Initial Equation Variable (START:).} The predicate variable listed after ``START:'' is the initial equation variable.
	\item \strong{Logical Formula (EQUATIONS:).} The formula after ``EQUATIONS:'' and within the braces is an alternation-free $L_{\nu,\mu}^{rel}$ formula. There are also a few implementation shortcuts used in the PES Formula syntax. Additionally, substitutions of state variables are allowed in addition to clock resets.
	\item \strong{Location Invariants (INVARIANT:)}. The ``INVARIANT'' Block lists all of the invariants. Each invariant is described by a subset of atomic propositions, then followed by a ``->'' and then followed by a clock constraint. There is one specification per line. For time to advance, all of the invariants must be true. However, this is often shortcut by vacuity: if the automaton is currently not in a state where the premise of the implication is true, the invariant is vacuously true.
	\item \strong{Automata Edges (TRANSITIONS:).} The ``TRANSITIONS:'' section lists all of the transitions of the timed automaton. Each edge of the timed automata is a transition. Each transition starts with the guard in parentheses, and then has an ``->'' to separate the guard from the transition to the new state. After the ``->'', the destination state is specified in parentheses, the clocks to be reset are specified in braces, and each transition ends with a semicolon.
\end{itemize}

We illustrate this with an example. Figure \ref{fig:lst1} is the PES for the LEADER benchmark with 4 processes, using the formula for category ``bs''.

\begin{figure}[!t]
\begin{lstlisting}
#define CPD 2
CLOCKS: {x1,x2,x3,x4}
CONTROL: {p1,p2,p3,p4, p}
PREDICATE: {X}
START: X
EQUATIONS: {
1: nu X = ((p1 == 0 && p2 == 0 && p3==0)
	||(p1 == 0 && p2 == 0 && p4==0)
	||(p1 == 0 && p3 == 0 && p4==0)
	||(p2 == 0 && p3 == 0 && p4==0)
) && \forall time(\AllAct(X))
}
INVARIANT:
	p1 == 0 && p==0 -> x1 <= CPD
	p2 == 0 && p==0 -> x2 <= CPD
	p3 == 0 && p==0 -> x3 <= CPD
	p4 == 0 && p==0 -> x4 <= CPD
TRANSITIONS:
	(p2 == 0 && p1 == 0, x2 <= CPD && x1 <= CPD)->(p2 = 1){x2, x1};
	(p3 == 0 && p1 == 0, x3 <= CPD && x1 <= CPD)->(p3 = 1){x3, x1};
	(p3 == 0 && p2 == 0, x3 <= CPD && x2 <= CPD)->(p3 = 2){x3, x2};
	(p4 == 0 && p1 == 0, x4 <= CPD && x1 <= CPD)->(p4 = 1){x4, x1};
	(p4 == 0 && p2 == 0, x4 <= CPD && x2 <= CPD)->(p4 = 2){x4, x2};
	(p4 == 0 && p3 == 0, x4 <= CPD && x3 <= CPD)->(p4 = 3){x4, x3};
	(p==0 && p1==0 && p2!=0 && p3!=0 && p4!=0)->(p=1){x1, x2, x3, x4};
\end{lstlisting}
\caption{Full PES Example used to illustrate the different components of a PES.}
\label{fig:lst1}
\end{figure}

\begin{remark}[Transitions are syntactic sugar] The PES framework is general enough that timed automata edges can either be specified within the PES formula (using substitutions and resets in addition to atomic propositions) or by using the TRANSITIONS directly. See Example \ref{ex:twoways}.

Performance can be greatly enhanced by using the ``TRANSITIONS'' feature to separate the model from the specification, but there is no requirement to do so.
\end{remark}

\begin{example}[Encoding a PES in multiple ways]
Consider timed automaton $TA_6$, given in \figref{fig:pesTA6}, with one clock $x$, and consider the following timed modal equation system $MES$ with one equation (and hence one block):
\begin{align}
X \mesnueq& \mmEXra{a}{X} 
\end{align}

\begin{figure}[ht]
\includegraphics[scale=0.7]{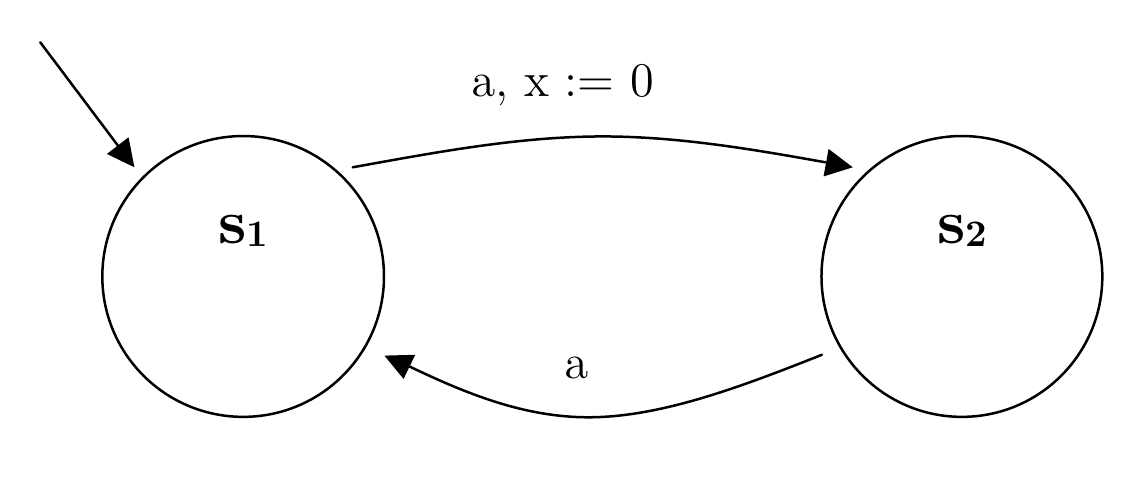}
\caption{Timed automaton $TA_6$.}
\label{fig:pesTA6}
\end{figure}

Note that this automaton has \strong{zeno} runs; we use it to provide a simple example on an untimed formula.

We encode this model into our tool using TRANSITIONS as follows:

\begin{lstlisting}
CLOCKS: {x1}
CONTROL: {p1}
INITIALLY: x1 == 0
PREDICATE: {X}
START: X
EQUATIONS: {
1: nu X =\ExistAct(X)	
}
TRANSITIONS:
	(p1 == 0)->(p1=1){x1};
	(p1 == 1)->(p1=0);
\end{lstlisting}

Alternatively, we can use the original Predicate Equation System (PES) form by encoding the transitions as proposition assignments with resets. In this case, we have two equivalent alternative versions

\begin{lstlisting}
CLOCKS: {x1}
CONTROL: {p1}
INITIALLY: x1 == 0
PREDICATE: {X}
START: X
EQUATIONS: {
1: nu X =((p1==0)->X[p1=1]{x1}) && ((p1==1)->X[p1=0])
}
\end{lstlisting}

or

\begin{lstlisting}
CLOCKS: {x1}
CONTROL: {p1}
INITIALLY: x1 == 0
PREDICATE: {X1, X2}
START: X1
EQUATIONS: {
1: nu X1 = X2{x1})
1: nu X2 = X1
}
\end{lstlisting}

Notice that we have two choices: represent location variables with new predicate variables, or to use new control variables. This equivalence gives the intuition behind the correctness of blocking and computing fixpoints simultaneously when their parities are the same (Bekic's Theorem, see \cite{bekic_definable_1984}). The ``1:'' is the equation block number, indicating that both equations are in the same block.

The Model checker would then ask the question: does the initial state $(s_1, \mset{x = 0})$ satisfy $X$?  (One can show that this timed automaton does satisfy the formula; the formal details are outside the scope of this document.)
\label{ex:twoways}
\end{example}

Some more implementation details specific to the PES are described in the following subsections.

\subsection{Equation Syntax}

Equations support the $L^{rel}_{\nu,\mu}$ syntax. Additionally, to aid in the efficiency of model checking, there are two formulas that are hard-coded shorthands for $L^{rel}_{\nu,\mu}$ subformulae. These are $UnableWaitInf$ and $AbleWaitInf$. They are:
\begin{align}
UnableWaitInf :=&\ \tfoEsh{\tfreeze{z}{\tfoAsh{z \leq 1}}} \\
AbleWaitInf :=&\ \tfoAsh{\tfreeze{z}{\tfoEsh{z > 1}}}
\end{align}

These two subformulas are very important in expressing liveness (AF) properties and their duals. See \cite{fontana_towards_2014} (Lemma 4.7.2) for more information. Although these formula utilize freeze quantification, the implementation of these can be implemented without subformulas and instead just checking if one can diverge in a state by examining the invariants.

\subsection{Invariant Syntax}

Invariants are always constructed as a set of conditions, with one line per condition. Each condition starts with a partial conjunction of state variable constraints, then has a ``->'' and then has a conjunction of clock constraints. The invariant is interpreted as the conjunction of conditions.

\subsection{Transition Syntax}

Transitions encode the variables checked for the start state, then the guard of clock constraints after a comma ``,'' symbol, then a ``->'', then the change of state variables to mark the destination state, and then in ``\{'' ``\}'' the set of clocks that are reset to 0 on the transition. Each transition is separated by a semicolon ``;''.

Consider this example transition:

\begin{verbatim}
(p5==2 && p==5, x5 > CB)->(p5=3, p=5);
\end{verbatim}

It adds a transition for all states with $p5 = 2$ and $p = 5$ and insists that clock $x5$ must have value at least the constant $CB$ (Defined above with a ``\#define CB 19''). When this transition executes, it changes the state value of $p5$ from $2$ to $3$ and maintains the value of $p$ at 5. It does not reset any clocks.

\section{CSMA/CD}

	The \emph{Carrier Sense, Multiple Access with Collision Detection} (CSMA/CD) example description is also taken from \cite{yovine-kronos:-a-1997}, which also provided figures that these figures are based on. Additionally, \cite{tanenbaum-computer-networks-2002} was consulted when generating these notes.

\subsection{Overview}

	There are multiple processes ($n$) who are sharing $1$ Ethernet bus.  The processes will wish to send messages, but the bus can only send one message at a time.  At various times processes will try to transmit a message.  If the process detects that the bus is busy (thus, the other process has transmitted for at least $\sigma$, the worst-case propagation delay, units), then the process will wait a random amount of time before it will retry without interrupting the currently transmitting process.  If multiple processes are transmitting, which would happen if they try to transmit simultaneously or in intervals before it can detect that the other processes are transmitting, a collision will occur, and all current messages are lost and must be sent again.  At this point the bus detects the collision and stops transmitting. Thus, even though only one process can send a message at a time, it is possible for more than one process to be in a transmit state (in these cases the bus will be in the collision state).  All processes then detect the collision and the bus goes from the collision state to the idle state.  With the collision detected, all transmitting processes retry, idle processes either remain idle or retry and processes programmed to retry reset their waiting cycle before they retry.
	Parameters:
	
\begin{itemize}
	\item $\mathbf{\boldsymbol{\sigma}}$: The worst case propagation delay.  The default value is $26\mu s$.  This is the worst case time for a packet to be detected. Thus, a process may not detect if another process is sending if that process has been sending for less than $\sigma$ units.
	\item $\mathbf{\boldsymbol{\lambda}}$: The time to send a complete message (including the propagation delay).  The default value is $808\mu s$.
\end{itemize}

\begin{remark}
This model assumes that at no time do two processes begin to transmit simultaneously.  A collision thus only occurs when one process is transmitting, and another tries to transmit without detecting it.
\end{remark}

\subsection{Timed Automata Modeling}
	
	These are the actions which the processes and the bus are ``synchronized'' on.  the $i$ subscript represents process $i$:
	
	\begin{itemize}
		\item $\texttt{begin}_{i}$: Process (Sender) $i$ starts sending a message.
		\item $\texttt{busy}_{i}$: Process $i$ senses that the bus is busy.
		\item $\texttt{end}_{i}$: Process $i$ ends the transmission of the message.
		\item $\texttt{cd}$: All processes detect a collision on a bus. 
		
		 All currently transmitting processes go to the retry state, those retrying retry again and each idle process either remains idle or decides to retry sending. 
	\end{itemize}
	
\begin{figure}
	\centering
	\includegraphics[scale=0.5]{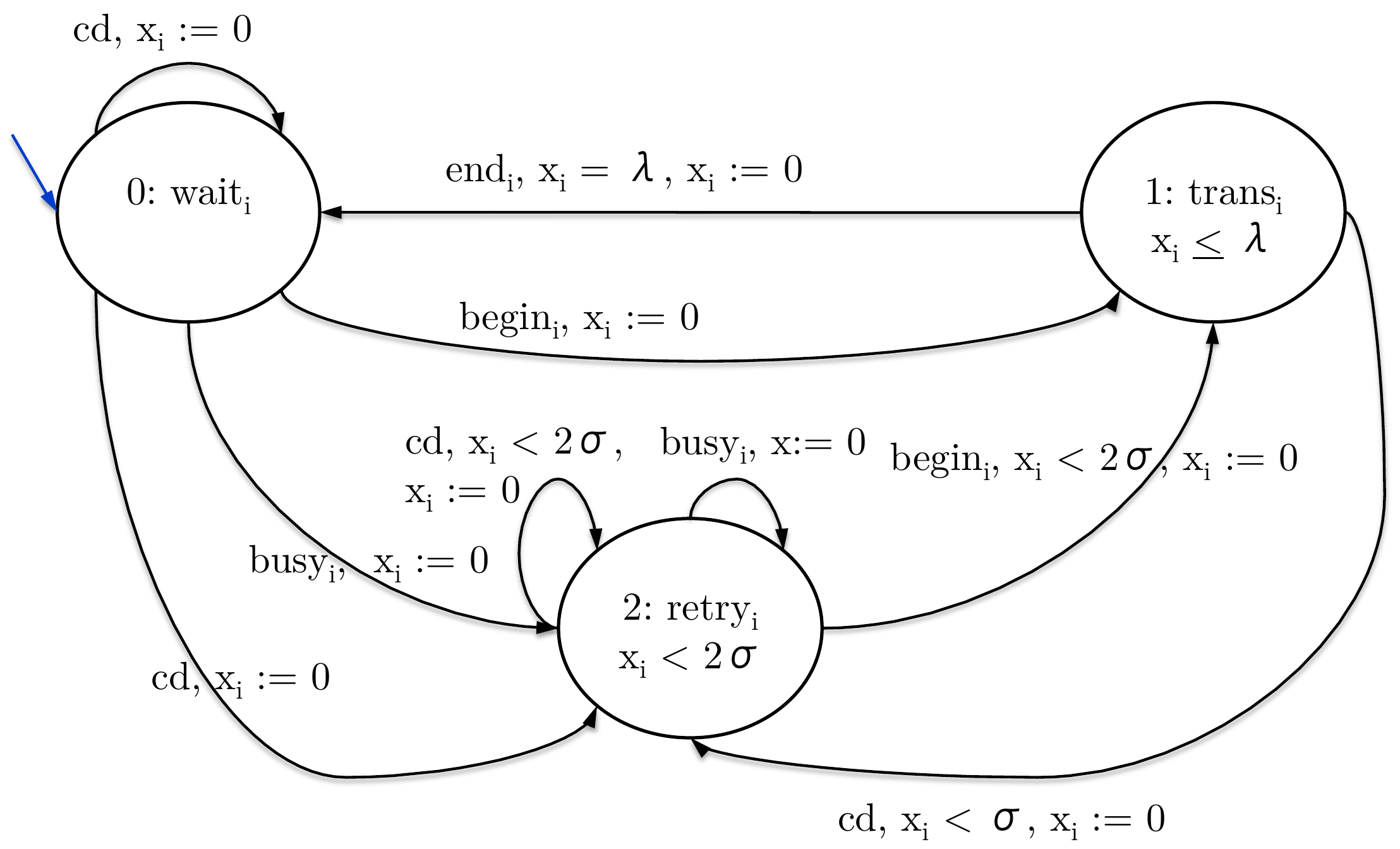}
	\caption{Timed Automaton for process $i$ in the CSMA/CD Example.}
	\label{fig:csmacdp}
	\end{figure}

	The Timed Automaton Figure for a single process (the $i^{th}$ process) is in Figure \ref{fig:csmacdp}
	
	The states of Process $i$ are as follows:
	
	\begin{enumerate}		
		\item[0.] $\texttt{wait}_{i}$: Process $i$ is waiting for a message.
		\item $\texttt{trans}_{i}$: Process $i$ is sending a message.
		\item $\texttt{retry}_{i}$: Process $i$ is waiting to retry (usually after detecting a  collision).
	\end{enumerate}
	
\begin{figure}
	\centering
	\includegraphics[scale=0.5]{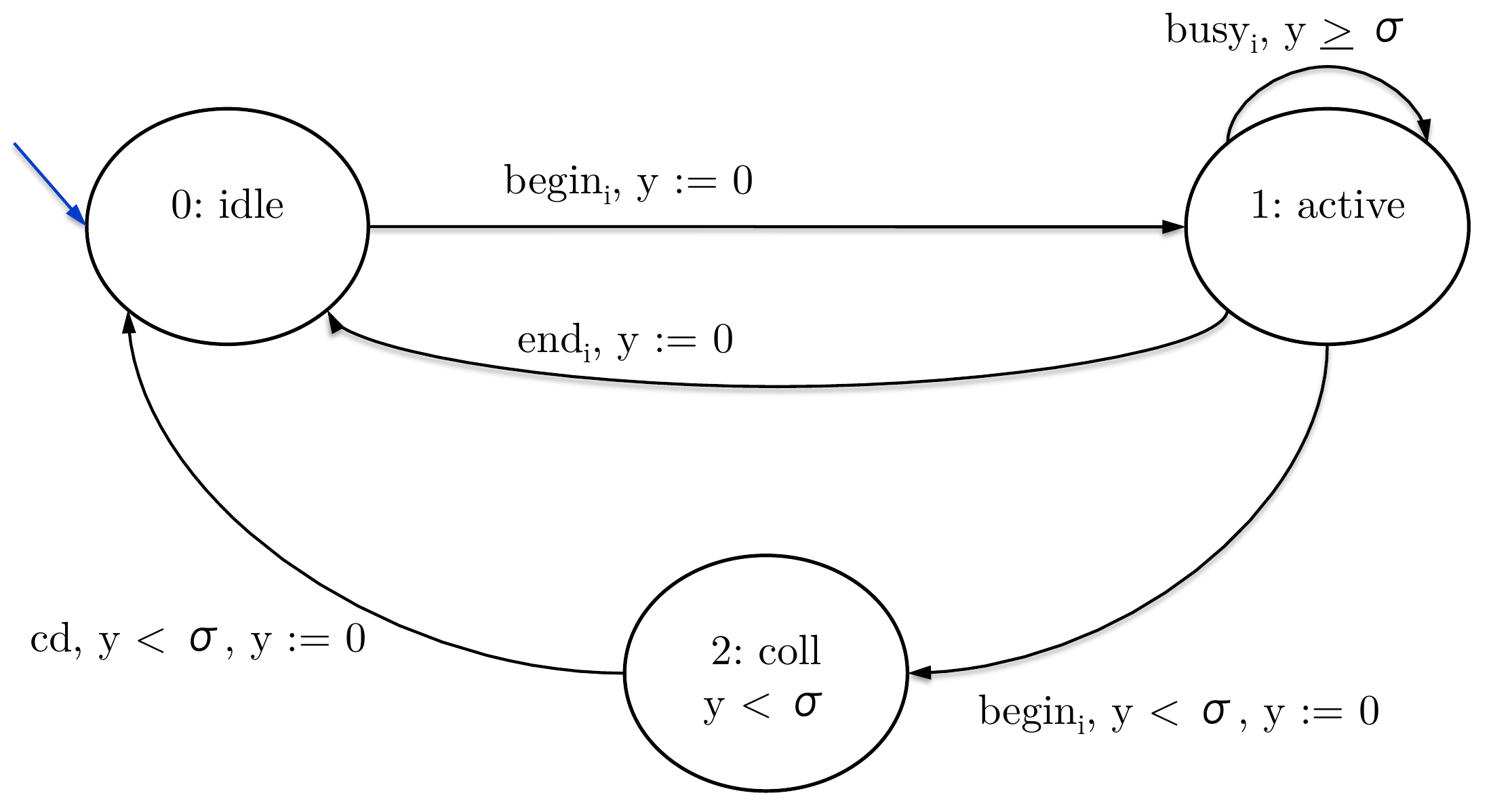}
	\caption{Timed Automaton for the Bus in the CSMA/CD Example.}
	\label{fig:csmacdb}
	\end{figure}

	The Timed Automaton Figure for the bus is in Figure \ref{fig:csmacdb}
	
	The States of the Bus are:
	
	\begin{enumerate}
		\item[0.] $\texttt{idle}$: The bus is currently idle and waiting for a message from a process.
		\item $\texttt{active}$: The bus is currently transmitting a message from some process.
		\item $\texttt{collision}$: There is a collision on the bus.
	\end{enumerate}
	
		For the notation of the Bus Automata, there is an transition with symbol $i$ for each process $i$.  Thus, the edge idle $\ttrans{\textrm{begin}_i}$ active$[y:=0]$ represents a different edge for each process.  If there are $2$ processes, then the bus would have $2$ such edges, $\ttrans{\textrm{begin}_1}$ active$[y:=0]$ and $\ttrans{\textrm{begin}_2}$ active$[y:=0]$.  Each collision detection involves a pair of precesses.  Thus it involves an event $\texttt{cd}_{ij}$ there is an edge per pair/trio of cd's).
	
	Each process has a clock $x_i$, and the bus has its own clock $y$.  Initially, we have $y = 0, x_i = 0$ (all $i$) and each process $i$ is in the initial state $0:wait_i$, and the bus is initially in $0:idle$.
	
\subsection{Specifications Checked}

The specifications checked on the CSMA protocol are: \parspc

\noindent \emph{\textbf{AS$^{*}$:}  At most one process is in a transmission state for less than 52 ($2\sigma$) units. (Valid)}

TCTL Formula: 
\begin{align} \displaystyle
\ctlAG{\bigwedge_{i\neq j}(p_i \neq 1 \lgcor p_j \neq 1) \lgcor (x_1 \leq 52 \lgcand x_2 \leq 52)}
\end{align}

The $L^{rel}_{\nu,\mu}$ is included in the Tool MES Specification (with abbreviations for notational convenience).

Tool MES Specification:
\begin{lstlisting}
#define CB       52
PREDICATE: {X}
START: X
EQUATIONS: {
1: nu X = \forall time((p1 != 1 || p2 != 1 || ((x1 < CB) && (x2 < CB))) && \AllAct(X))
}
\end{lstlisting} \parspc

\noindent \textbf{BS:} At any time, a third process can retry while two are already in transmission status. (Invalid)

TCTL Formula: 
\begin{align} \displaystyle
\ctlAG{\bigwedge_{i \neq j \neq k \lgcand i \neq k}\Bigl((p_i = 1 \lgcand p_j = 1) \lgcif p_k = 2)\Bigr)}
\end{align}

The $L^{rel}_{\nu,\mu}$ is included in the Tool MES Specification (with abbreviations for notational convenience).

Tool MES Specification:
\begin{lstlisting}
PREDICATE: {X}
START: X
EQUATIONS: {
1: nu X =  (((p1==1 && p2==1)->(p3==2)) && ((p1==1 && p3==1)->(p2==2)) && ((p2==1 && p3==1)->(p1==2)))
 && \forall time(\AllAct(X))
}
\end{lstlisting} \parspc

\noindent \textbf{AL:}  It is inevitable that all processes are waiting. (Valid)

TCTL Formula: 
\begin{align} \displaystyle
\ctlAF{\bigwedge_{i}\Bigl(p_i = 0 \lgcand x_i \geq 52\Bigr)}
\end{align}

The $L^{rel}_{\nu,\mu}$ is included in the Tool MES Specification (with abbreviations for notational convenience).

Tool MES Specification:
\begin{lstlisting}
#define CB       52
PREDICATE: {X}
START: X
EQUATIONS: {
1: mu X = \forall time\rel[((p1 == 0 && x1 >= CB) || (p2 == 0 && x2 >= CB))](((p1 == 0 && x1 >= CB) || (p2 == 0 && x2 >= CB)) || \AllAct(X)) && (UnableWaitInf || \exists time(((p1 == 0 && x1 >= CB) || (p2 == 0 && x2 >= CB))))
}
\end{lstlisting} \parspc

\noindent \textbf{BL:} It is inevitable that some process needs to retry transmitting a message. (Invalid)

TCTL Formula: 
\begin{align} \displaystyle
\ctlAF{\bigvee_{i}\Bigl(p_i = 2\Bigr)}
\end{align}

The $L^{rel}_{\nu,\mu}$ is included in the Tool MES Specification (with abbreviations for notational convenience).

Tool MES Specification:
\begin{lstlisting}
PREDICATE: {X}
START: X
EQUATIONS: {
1: mu X = \forall time\rel[((p1 == 2) || (p2 == 2))](((p1 == 2) || (p2 == 2)) || \AllAct(X)) && (UnableWaitInf || \exists time(((p1 == 2) || (p2 == 2))))
}
\end{lstlisting} \parspc

\noindent \textbf{M1:} It is always the case that if the first process needs to retry that it will inevitably transmit. (Invalid) 

TCTL Formula: 
\begin{align} \displaystyle
\ctlAG{p_1 \neq 2 \lgcif \ctlAF{p_1 = 1}}
\end{align}

The $L^{rel}_{\nu,\mu}$ is included in the Tool MES Specification (with abbreviations for notational convenience).

Tool MES Specification:
\begin{lstlisting}
PREDICATE: {X,X2}
START: X
EQUATIONS: {
1: nu X = \forall time( ({p1 != 2} || X2) && \AllAct(X))
2: mu X2 = \forall time\rel[(p1 == 1)]((p1 == 1) || \AllAct(X2)) && (UnableWaitInf || \exists time((p1 == 1)))
}
\end{lstlisting} \parspc

\noindent \textbf{M2:} It is always the case that if a bus experiences a collision that it will inevitably become idle. (Valid)

TCTL Formula: 
\begin{align} \displaystyle
\ctlAG{p \neq 2 \lgcif \ctlAF{p = 0}}
\end{align}

The $L^{rel}_{\nu,\mu}$ is included in the Tool MES Specification (with abbreviations for notational convenience).

Tool MES Specification:
\begin{lstlisting}
PREDICATE: {X,X2}
START: X
EQUATIONS: {
1: nu X = \forall time( ({p != 2} || X2) && \AllAct(X))
2: mu X2 = \forall time\rel[(p == 0)]((p == 0) || \AllAct(X2)) && (UnableWaitInf || \exists time((p == 0)))
}
\end{lstlisting} \parspc

\noindent \emph{\textbf{M3$^{*}$:} The bus is always idle until a process is active. (Invalid)} 

TCTL Formula: 
\begin{align} \displaystyle
\ctlAU{p = 0}{\bigvee_{i}\Bigl(p_i = 1\Bigr)}
\end{align}

The $L^{rel}_{\nu,\mu}$ is included in the Tool MES Specification (with abbreviations for notational convenience).

Tool MES Specification:
\begin{lstlisting}
PREDICATE: {X}
START: X
EQUATIONS: {
1: mu X = \forall time\rel[((p1 == 1) || (p2 == 1))]( ((p == 0) || ((p1 == 1) || (p2 == 1))) && (((p1 == 1) || (p2 == 1)) || \AllAct(X))) && ( UnableWaitInf || \exists time\rel[(p == 0)](((p1 == 1) || (p2 == 1))))}
\end{lstlisting} \parspc

\noindent \emph{\textbf{M4$^{*}$:} For all paths with an infinite number of actions, the bus is always idle until a process is active (Valid)}

TCTL Formula: 
\begin{equation}
\text{There is no known TCTL formula}\nonumber
\end{equation}

The $L^{rel}_{\nu,\mu}$ is included in the Tool MES Specification (with abbreviations for notational convenience).

Tool MES Specification:
\begin{lstlisting}
PREDICATE: {X}
START: X
EQUATIONS: {
1: mu X = ((p1 == 1) || (p2 == 1)) || ((p == 0) && \forall time(\AllAct(X)))
}
\end{lstlisting} \parspc

\subsection{Model in PES}

Below is the CSMA/CD Model for 2 processes specified in PES Code. The equation is the formula, and for this model the formula is the ``as'' formula.

\begin{lstlisting}
#define CA       26
#define CB       52
#define CLAMBDA  808
CLOCKS: {x1,x2,y}
CONTROL: {p1,p2,p}
PREDICATE: {X}
START: X
EQUATIONS: {
1: nu X = \forall time((p1 != 1 || p2 != 1 || ((x1 < CB) && (x2 < CB))) && \AllAct(X))
}
INVARIANT:
	p1 == 1 -> x1 <= CLAMBDA
	p1 == 2 -> x1 < CB
	p2 == 1 -> x2 <= CLAMBDA
	p2 == 2 -> x2 < CB
	p == 2 -> y < CA
TRANSITIONS:
	(p==0 && p1==0)->(p=1,p1=1){y,x1};
	(p==0 && p1==2, x1 < CB)->(p=1,p1=1){y,x1};
	(p==0 && p2==0)->(p=1,p2=1){y,x2};
	(p==0 && p2==2, x2 < CB)->(p=1,p2=1){y,x2};
	(p==1 && p1==1, x1 == CLAMBDA)->(p=0,p1=0){y,x1};
	(p==1 && p2==1, x2 == CLAMBDA)->(p=0,p2=0){y,x2};
	(p==1 && p1==0, y >= CA)->(p=1,p1=2){x1};
	(p==1 && p1==2, y >= CA)->(p=1,p1=2){x1};
	(p==1 && p2==0, y >= CA)->(p=1,p2=2){x2};
	(p==1 && p2==2, y >= CA)->(p=1,p2=2){x2};
	(p==1 && p1==0, y < CA)->(p=2,p1=1){y,x1};
	(p==1 && p1==2, y < CA && x1 < CB)->(p=2,p1=1){y,x1};
	(p==1 && p2==0, y < CA)->(p=2,p2=1){y,x2};
	(p==1 && p2==2, y < CA && x2 < CB)->(p=2,p2=1){y,x2};
	(p == 2 && p1==0 && p2==0, y < CA)->(p=0,p1=0,p2=0){y};
	(p == 2 && p1==0 && p2==0, y < CA)->(p=0,p1=0,p2=2){y,x2};
	(p == 2 && p1==0 && p2==1, y < CA && x2 < CA)->(p=0,p1=0,p2=2){y,x2};
	(p == 2 && p1==0 && p2==2, y < CA && x2 < CB)->(p=0,p1=0,p2=2){y,x2};
	(p == 2 && p1==0 && p2==0, y < CA)->(p=0,p1=2,p2=0){y,x1};
	(p == 2 && p1==0 && p2==0, y < CA)->(p=0,p1=2,p2=2){y,x1,x2};
	(p == 2 && p1==0 && p2==1, y < CA && x2 < CA)->(p=0,p1=2,p2=2){y,x1,x2};
	(p == 2 && p1==0 && p2==2, y < CA && x2 < CB)->(p=0,p1=2,p2=2){y,x1,x2};
	(p == 2 && p1==1 && p2==0, y < CA && x1 < CA)->(p=0,p1=2,p2=0){y,x1};
	(p == 2 && p1==1 && p2==0, y < CA && x1 < CA)->(p=0,p1=2,p2=2){y,x1,x2};
	(p == 2 && p1==1 && p2==1, y < CA && x1 < CA && x2 < CA)->(p=0,p1=2,p2=2){y,x1,x2};
	(p == 2 && p1==1 && p2==2, y < CA && x1 < CA && x2 < CB)->(p=0,p1=2,p2=2){y,x1,x2};
	(p == 2 && p1==2 && p2==0, y < CA && x1 < CB)->(p=0,p1=2,p2=0){y,x1};
	(p == 2 && p1==2 && p2==0, y < CA && x1 < CB)->(p=0,p1=2,p2=2){y,x1,x2};
	(p == 2 && p1==2 && p2==1, y < CA && x1 < CB && x2 < CA)->(p=0,p1=2,p2=2){y,x1,x2};
	(p == 2 && p1==2 && p2==2, y < CA && x1 < CB && x2 < CB)->(p=0,p1=2,p2=2){y,x1,x2};
\end{lstlisting}

\section{FISCHER's MUX}

	The \emph{Fischer's Mutual Exclusion} (MUX) example description is also taken from \cite{balarin-approximate-reachability-1996}, \cite{behrmann-a-tutorial-2004}, \cite{alur-an-implementation-1992} and \cite{daws-the-tool-1996}, which also provided figures that these figures are based on.

\subsection{Overview}

	The FISCHER MUX Protocol involves $n$ processes vying for access to a critical section.  The Critical Section is modeled as a state machine representing the process that currently has access to the critical section or if the critical section is available.  Each process asks for the critical section and then waits until it gets it, re-requesting for access if it is not granted it for a period of time. The critical section represents a global variable of which process currently has access to the critical section (like a lock).  Each process assigns the variable to itself when requesting to enter the critical section and sets it to $0$ when they have left the critical section.  The value of the variable keeps track of whose turn it is to access the critical section (process in state $3$).  Note that a process can assign the variable to itself and not yet be in the critical section.
		
	Parameters:
	
\begin{itemize}
	\item $\mathbf{\Delta}$ ($CA$): This is the time the process must delay before it assigns the global variable to itself, thus vying for the critical section.  The default value is $10$.
	\item $\mathbf{\boldsymbol\delta}$ ($CB$): This gives the amount of time a process must wait before entering the critical section after claiming access to it.  The default value is $19$.
\end{itemize}

\subsection{Timed Automata Modeling}
	
	These are the actions which the processes and the global section are ``synchronized'' on. 	
	\begin{itemize}
		\item	$\mathbf{try_i}$ Here process $i$ will try to access the critical section
		\item $\mathbf{setX_i}$ Here the process sets the global variable to value $i$.
		\item $\mathbf{setX0_i}$ Here process $i$ sets the global variable to $0$ (section empty).
		\item $\mathbf{enter_i}$  Here process $i$ enters the critical section.
	\end{itemize}
	
\begin{figure}
	\centering
	\includegraphics[scale=0.5]{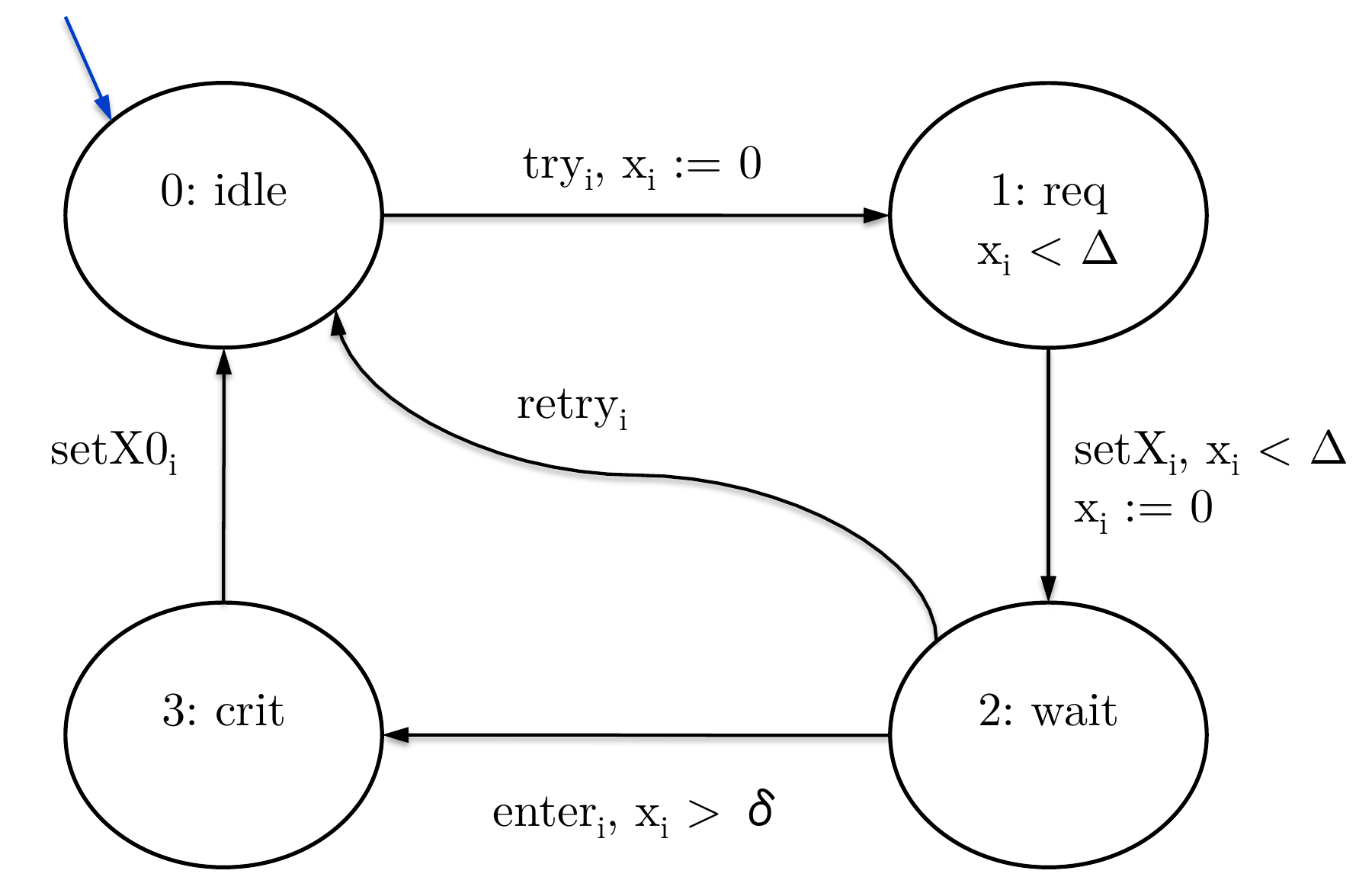}
	\caption{Timed Automaton for a Process in the MUX Example.}
	\label{fig:fischerp}
	\end{figure}

	The Timed Automaton Figure for a single process is in Figure \ref{fig:fischerp}.
	
	The states of Process $i$ are as follows:
	
	\begin{enumerate}		
		\item [0.] $\texttt{idle:}$ The process is idle.
		\item [1.] $\texttt{req:}$ The process is ready to enter the critical section and will ask to do so
		\item [2.] $\texttt{wait:}$ The process is waiting to enter the critical section
		\item [3.] $\texttt{crit:}$ The process is in the critical section.	
	\end{enumerate}

\begin{remark}
	Here we follow  \cite{wang-efficient-verification-2003} and \cite{balarin-approximate-reachability-1996} and include the edge given from state $2$ to state $0$ to prevent starvation.  This edge is omitted in many sources, including \cite{behrmann-a-tutorial-2004}, \cite{alur-an-implementation-1992} and \cite{daws-the-tool-1996}.  Some of those sources, instead have an edge $(2, try_{i}, x_i > \delta, \mset{x_i := 0}, 1)$.  That edge from state $2$ to state $1$ is given to prevent \emph{starvation}, so that every process will be able to access the critical section eventually.  Some models (including \cite{alur-an-implementation-1992}) omit this edge as well.  Also some models omit the invariant on state $1$, which prevents a timelock.
\end{remark}

\begin{figure}
	\centering
	\includegraphics[scale=0.5]{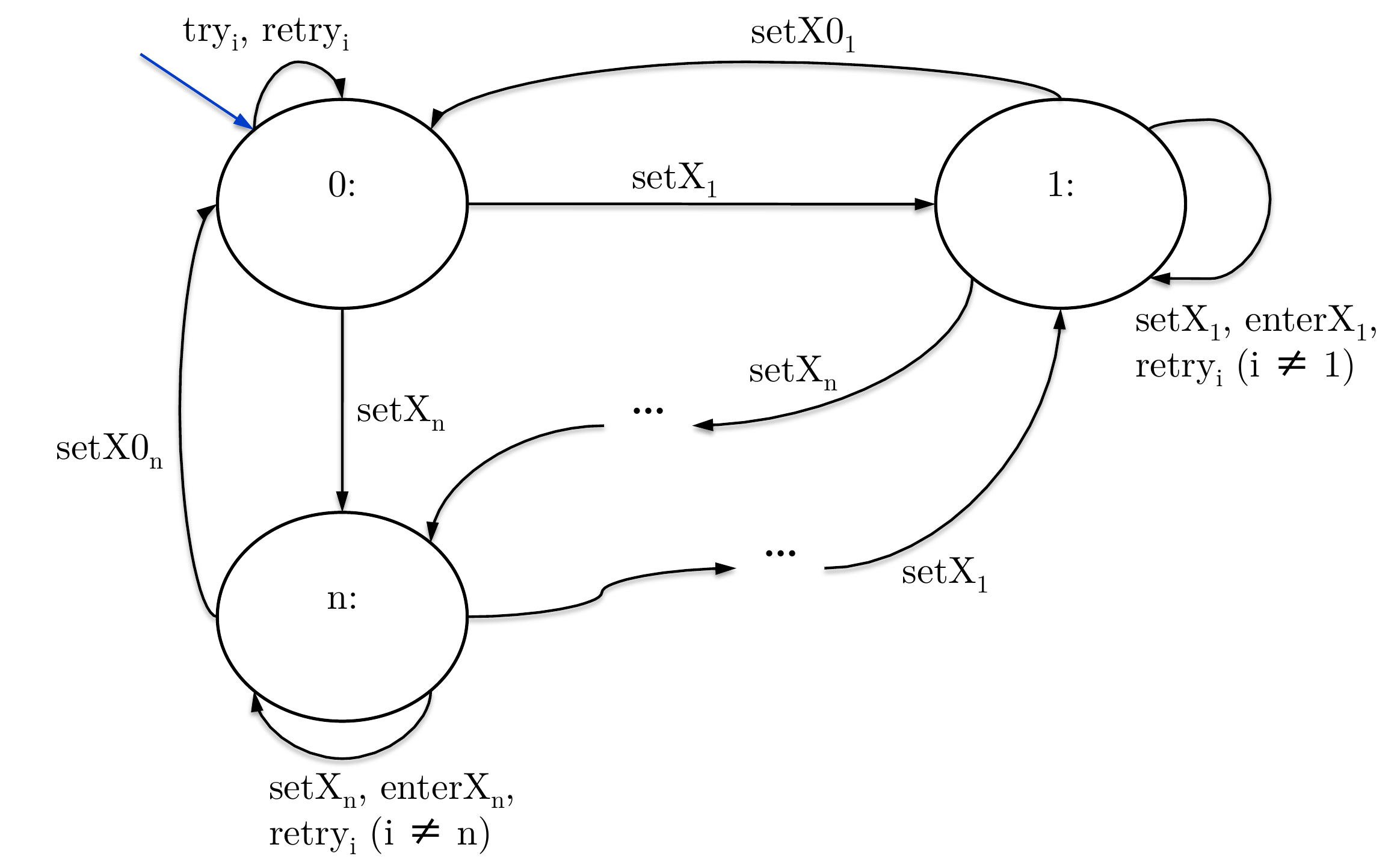}
	\caption{Timed Automaton for the Critical Section in the MUX Example.}
	\label{fig:fischerc}
	\end{figure}

	The Timed Automaton Figure for the Critical Section is in Figure \ref{fig:fischerc}
	
	There are $n + 1$ states of the Critical Section.  They are:
	
	\begin{enumerate}
		\item State $0$.  The Critical Section is empty.		
		\item States 1 through $n$. The Critical Section is being used by processor $i$ (processor $1$ is the first processor).
	\end{enumerate}
	
	Notice that the more processes there are, the more states the automata has.  Here, the token is passed around in a cyclic manner, where
		
	Each process has a clock $x_i$, while the global (critical) section has no clock.  Initially, each clock is 0 $x_i = 0$ (for all i), each process is initially in state $0:idle$ and the critical section is initially in state $0$ (empty).

\subsection{Specifications Checked}

The specifications checked on the FISCHER protocol are:

\noindent \textbf{AS:}  At any time, at most one process is in the critical section. (Valid)

TCTL Formula: 
\begin{align} \displaystyle
\ctlAG{\bigwedge_{i\neq j}(p_i \neq 3 \lgcor p_j \neq 3)}
\end{align}

The $L^{rel}_{\nu,\mu}$ is included in the Tool MES Specification (with abbreviations for notational convenience).

Tool MES Specification:
\begin{lstlisting}
PREDICATE: {X}
START: X
EQUATIONS: {
1: nu X = (p1 != 3 || p2 != 3) && \forall time(\AllAct(X))
}
\end{lstlisting} \parspc

\noindent \textbf{BS:} At any moment, at most four processes in their waiting state at the same time. (Valid for four processes, Invalid for five or more processes)

TCTL Formula: 
\begin{align} \displaystyle
\ctlAG{p_1 \neq 5)}
\end{align}

The $L^{rel}_{\nu,\mu}$ is included in the Tool MES Specification (with abbreviations for notational convenience).

Tool MES Specification:
\begin{lstlisting}
PREDICATE: {X}
START: X
EQUATIONS: {
1: nu X = (p1 != 5)&& \forall time(\AllAct(X))
}
\end{lstlisting} \parspc

\noindent \textbf{AL:} It is inevitable that all processes are idle. (Valid)

TCTL Formula: 
\begin{align} \displaystyle
\ctlAF{\bigwedge_{i\neq j}(p_i = 0 \lgcor p_j = 0}
\end{align}

The $L^{rel}_{\nu,\mu}$ is included in the Tool MES Specification (with abbreviations for notational convenience).

Tool MES Specification:
\begin{lstlisting}
PREDICATE: {X}
START: X
EQUATIONS: {
1: mu X = \forall time\rel[((p1 == 0) && (p2 == 0))](((p1 == 0) && (p2 == 0)) || \AllAct(X)) && (UnableWaitInf || \exists time(((p1 == 0) && (p2 == 0))))
}
\end{lstlisting} \parspc

\noindent \textbf{BL:} It is inevitable that some process accesses the critical section. (Invalid)

TCTL Formula: 
\begin{align} \displaystyle
\ctlAF{\bigvee_{i}(p_i = 2)}
\end{align}

The $L^{rel}_{\nu,\mu}$ is included in the Tool MES Specification (with abbreviations for notational convenience).

Tool MES Specification:
\begin{lstlisting}
PREDICATE: {X}
START: X
EQUATIONS: {
1: mu X = \forall time\rel[((p1 == 2) || (p2 == 2))](((p1 == 2) || (p2 == 2)) || \AllAct(X)) && (UnableWaitInf || \exists time(((p1 == 2) || (p2 == 2))))
}
\end{lstlisting} \parspc

\textbf{M1:} It is always the case that if the first process is not idle, it will eventually access the critical section. (Invalid)

TCTL Formula: 
\begin{align} \displaystyle
\ctlAG{p_1 \neq 0 \lgcif \ctlAF{p_1 = 3}}
\end{align}

The $L^{rel}_{\nu,\mu}$ is included in the Tool MES Specification (with abbreviations for notational convenience).

Tool MES Specification:
\begin{lstlisting}
PREDICATE: {X,X2}
START: X
EQUATIONS: {
1: nu X = \forall time( ({p1 == 0} || X2) && \AllAct(X))
2: mu X2 = \forall time\rel[(p1 == 3)]((p1 == 3) || \AllAct(X2)) && (UnableWaitInf || \exists time((p1 == 3)))
}
\end{lstlisting} \parspc

\textbf{M2:} It is always the case that if the third process is not idle, it will eventually access the critical section. (Invalid)

TCTL Formula: 
\begin{align} \displaystyle
\ctlAG{p_3 \neq 0 \lgcif \ctlAF{p_3 = 3}}
\end{align}

The $L^{rel}_{\nu,\mu}$ is included in the Tool MES Specification (with abbreviations for notational convenience).

Tool MES Specification:
\begin{lstlisting}
PREDICATE: {X,X2}
START: X
EQUATIONS: {
1: nu X = \forall time( ({p3 == 0} || X2) && \AllAct(X))
2: mu X2 = \forall time\rel[(p3 == 3)]((p3 == 3) || \AllAct(X2)) && (UnableWaitInf || \exists time((p3 == 3)))
}
\end{lstlisting} \parspc

\emph{\textbf{M3$^{*}$:} It is possible for the first process to enter the critical section without waiting. (Invalid)}

TCTL Formula: 
\begin{align} \displaystyle
\text{No Known TCTL Formula.}
\end{align}

The $L^{rel}_{\nu,\mu}$ is included in the Tool MES Specification (with abbreviations for notational convenience).

Tool MES Specification:
\begin{lstlisting}
PREDICATE: {X}
START: X
EQUATIONS: {
1: mu X = \exists time( (p1 == 3 && x1 <= 0) || \ExistAct(X))
}
\end{lstlisting} \parspc

\emph{\textbf{M4$^{*}$:} After at most five action transitions, some process will enter the critical section. (Invalid)}

TCTL Formula: 
\begin{align} \displaystyle
\text{No Known TCTL Formula.}
\end{align}

The $L^{rel}_{\nu,\mu}$ is included in the Tool MES Specification (with abbreviations for notational convenience).

Tool MES Specification:
\begin{lstlisting}
PREDICATE: {X}
START: X
EQUATIONS: {
1: nu X = \forall time( ((p1 == 3) || (p2 == 3)) || \AllAct(\forall time( ((p1 == 3) || (p2 == 3)) || \AllAct(\forall time( ((p1 == 3) || (p2 == 3)) || \AllAct(\forall time( ((p1 == 3) || (p2 == 3)) || \AllAct(\forall time( ((p1 == 3) || (p2 == 3)) || \AllAct(((p1 == 3) || (p2 == 3)))) )) )) )) ))
}
\end{lstlisting} \parspc

\subsection{Model in PES}

Below is the FISCHER Model for 2 processes specified in PES Code. The equation is the formula, and for this model the formula is the ``as'' formula.

\begin{lstlisting}
#define CA 10
#define CB 19
CLOCKS: {x1,x2}
CONTROL: {p1,p2,p}
PREDICATE: {X}
START: X
EQUATIONS: {
1: nu X = (p1 != 3 || p2 != 3) && \forall time(\AllAct(X))
}
INVARIANT:
	p1 == 1 -> x1 < CA
	p2 == 1 -> x2 < CA
TRANSITIONS:
	(p1==0 && p==0)->(p1=1, p=0){x1};
	(p1==1, x1 < CA)->(p1=2, p=1){x1};
	(p1==2 && p==1, x1 > CB)->(p1=3, p=1);
	(p1==2 && p!=1)->(p1=0);
	(p1==3)->(p1=0, p=0);
	(p2==0 && p==0)->(p2=1, p=0){x2};
	(p2==1, x2 < CA)->(p2=2, p=2){x2};
	(p2==2 && p==2, x2 > CB)->(p2=3, p=2);
	(p2==2 && p!=2)->(p2=0);
	(p2==3)->(p2=0, p=0);
\end{lstlisting}

\section{GRC}

	The \emph{Generalized Railroad Crossing} (GRC)example description is also taken from \cite{wang-symbolic-parametric-2004, zhang-fast-on-the-fly-2005, archer-mechanical-verification-1996, heitmeyer-a-benchmark-1993, heitmeyer-the-generalized-1994} and \cite{alur-an-implementation-1992}, which \cite{alur-an-implementation-1992} also provided figures that influenced these figures. are based on.  Unlike the other examples, this was not taken from \cite{zhang-fast-generic-2005}.  These figures are based off of various models and specifications, but is slightly different than all of them.
	
\subsection{Overview}

	The GRC (\emph{Generalized Railroad Crossing} Protocol involves $n$ trains (processess) who are each on a separate track that may want to cross through a region of interest.  This region is supervised by a gate that must be down when a train is crossing through and should be up when no train is in the region, and a controller that issues the gate instructions.	  
	
	The problem formulation is described well by \cite{heitmeyer-a-benchmark-1993}, quoted here:
	
	\begin{quotation}
	The system to be developed operates a gate at a railroad crossing. The railroad crossing $I$ lies in a region of interest $R$, i.e., $I \subseteq R$. A set of trains travel through $R$ on multiple tracks
in both directions. A sensor system determines when each train enters and exits region $R$. To describe the system formally, we define a gate function $g(t) \in [0, 90]$, where $g(t) = 0$
means the gate is down and $g(t) = 90$ means the gate is up. We also define a set \mset{\lambda_{i}} of \emph{occupancy intervals}, where each occupancy interval is a time interval during which one or
more trains are in $I$. The $i$th occupancy interval is represented as $\lambda_{i} = [\tau_{i}, \nu_{i}]$, where $\tau_i$ is
the time of the $i$th entry of a train into the crossing when no other train is in the crossing and $\nu_{i}$ is the first time since $\tau_{i}$ that no train is in the crossing (i.e., the train that entered at $\tau_{i}$ has exited as have any trains that entered the crossing after $\tau_{i}$).

Given two constants $\xi_1$ and $\xi_2$, $\xi_1 > 0, \xi_2 > 0$, the problem is to develop a system to operate the crossing gate that satisfies the following two properties:

\begin{description}
\item[\textbf{Safety Property}]  $t \in \cup_{i}\lambda_i \Rightarrow g(t) = 0$ (The gate is down during all occupancy intervals)

\item[\textbf{Utility Property}] $t \not\in \cup_{i}[\tau_i - \xi_1, \nu_i + \xi_2] \Rightarrow g(t) = 90$ (The gate is up as often
as possible)
\end{description}
	\end{quotation}

For our modeling, it involves $n$ trains that wish to cross.  When each train is approaching the track, it sends a signal to the controller to lower the gate (if the gate is not already down).  When it leaves, it sends a signal to the controller that it leaves.  The controller has a counter of the current number of trains that are approaching or in, so it knows when the first train enters and when the last one leaves.  A train is the ``last train'' if no other train is either near or in (requires a gate lower or down). After some delay, the controller can send a signal to raise or lower the gate.  The gate takes some time to raise or lower the gate.  If the gate is currently being raised when there is a signal to lower, it will switch directions, and is modeled to take the full amount of time to raise/lower the gate regardless of when it was switched midway.

\begin{remark}
Farn Wang in \cite{wang-symbolic-parametric-2004} used a linear hybrid automata to model the gate.  One condition we relax is if a signal to lower the gate is received by the gate while it is raising the gate (or vice versa), the gate takes the full amount of time to lower the gate as if it was already raised.  This is weaker than in \cite{wang-symbolic-parametric-2004} where the gate can start lowering from its current position and thus finish sooner.
\end{remark}

Parameters:
	
\begin{itemize}
	\item $\mathbf{CD}$ (\emph{Controller Delay}).  This parameter specifies the time that it takes for the controller to send a signal to the gate to raise/lower upon receiving a signal from a train.  (It is assumed that once the controller sends a signal, that the gate instantaneously receives it and responds, so this delay incorporates those delays).  The default value is 1.
	\item $\mathbf{GLT}$ (\emph{Gate Lowering Time})  This is the time it takes to lower a gate from the up position to the down position.  The default value is 2.
	\item $\mathbf{GRT}$ (\emph{Gate Raising Time})  This is the time it takes to raise a gate from the down position to the up position.  The default value is  2, (the same as $GRT$).
	\item $\mathbf{TP}$ (\emph{Train Pending}).  This is the amount of time in advance a train sends a signal to the controller that it is approaching the track.  The default value is 4.
	\item $\mathbf{TDU}$ (\emph{Train Duration Upper}).  This is an upper bound of time that the train is in the tracks.  The default value is 15.
	\item $\mathbf{TDL}$ (\emph{Train Duration Lower}).  This is a lower bound of time that the train is in the tracks.  The default value is 1.
\end{itemize}

\subsection{Timed Automata Modeling}

	These figures (as well as the symbols and state names) were created by us but based off of \cite{alur-an-implementation-1992}.
	
	These are the actions which the processes and the global section are ``synchronized'' on. 	
	\begin{itemize}
		\item \texttt{approach:} This is the signal the train gives to the controller when it is approaching the region.
		\item \texttt{exit:}  This is the signal the train gives to the controller when it leaves the region.
		\item \texttt{lower:}  This is the signal to begin lowering the gate.
		\item \texttt{raise:} This is the signal to begin raising the gate.
		\item \texttt{in:} This is the signal the train gives when it just enters the region.
		\item \texttt{down:} This is the signal the gate gives when it has finished lowering the gate (the gate is now down).
		\item \texttt{up:} This is the signal the gate gives when it has finished raising the gate (the gate is now up).
	\end{itemize}
	
\begin{figure}
	\centering
	\includegraphics[scale=0.5]{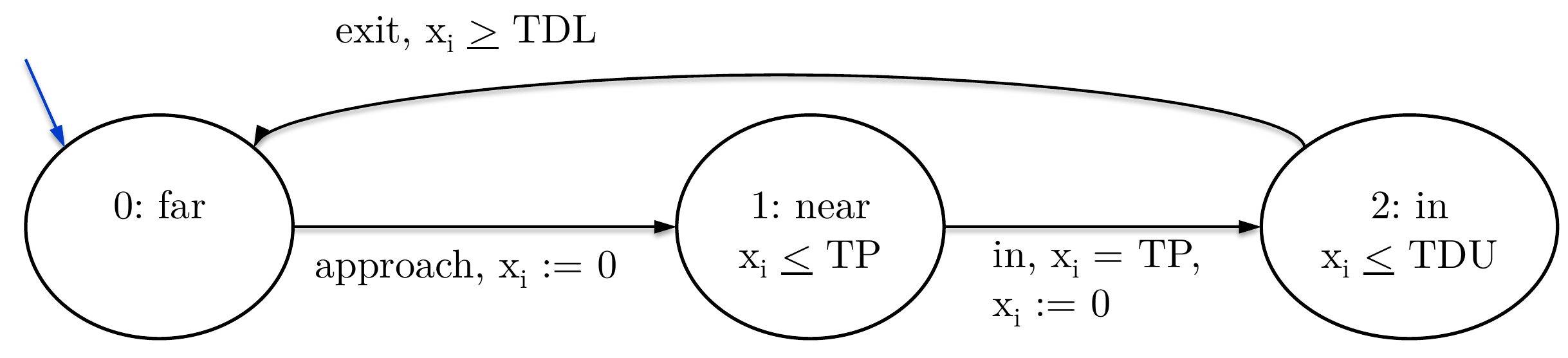}
	\caption{Timed Automaton for a Process (train) in the GRC Example.}
	\label{fig:grcp}
	\end{figure}

	The Timed Automaton Figure for a single process (train) is in Figure \ref{fig:grcp}.  The process's state is the parallel composition of the trains, gate and controller. 
		
	The states of Process (Train) $i$ are as follows:
	
	\begin{enumerate}		
		\item [0.] \texttt{far:} The train is far away.
		\item [1.] \texttt{near:} The train is approaching and will enter soon.		
		\item [2.] \texttt{in:} The train is currently in the region.
	\end{enumerate}

\begin{figure}
	\centering
	\includegraphics[scale=0.5]{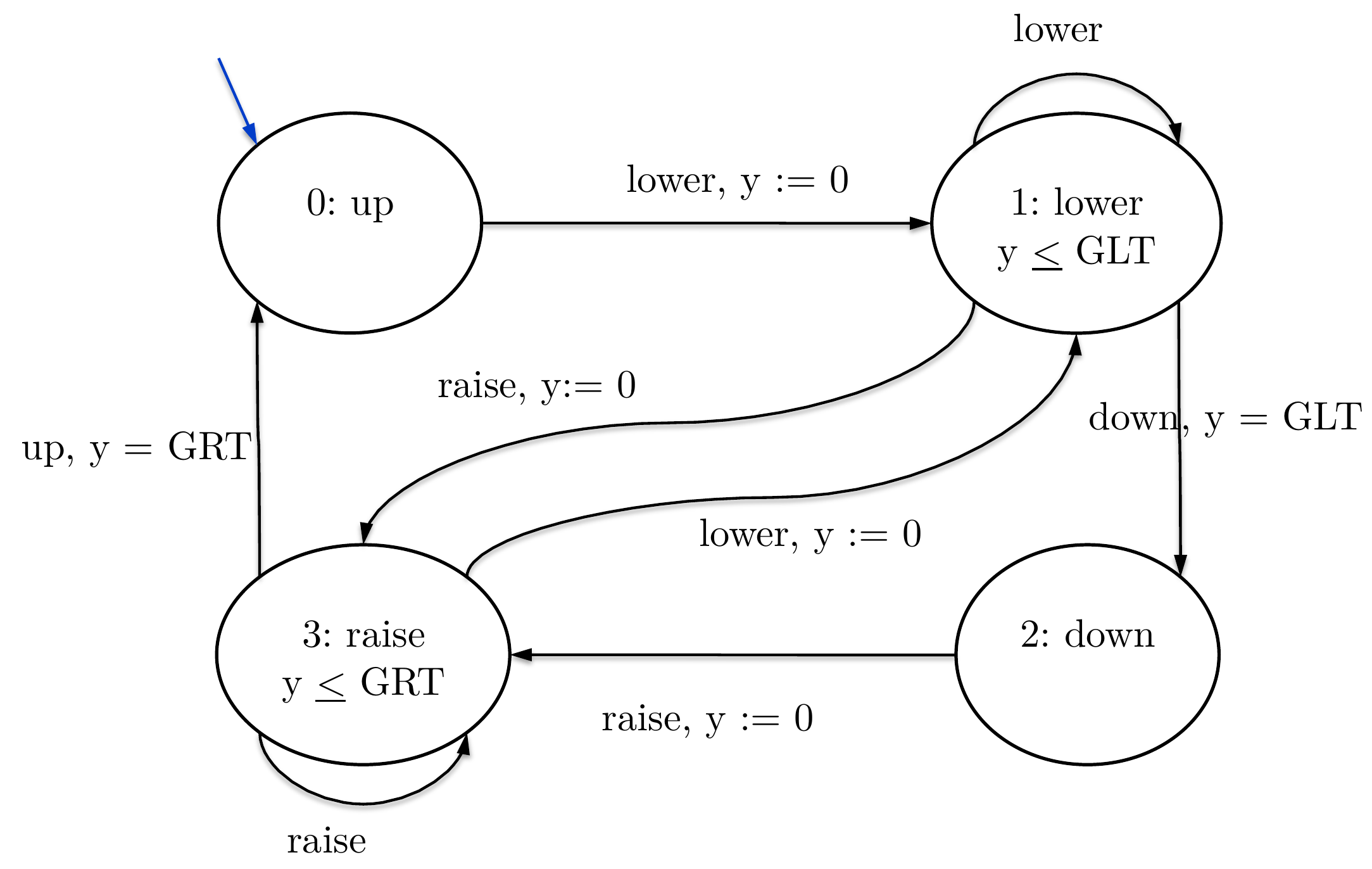}
	\caption{Timed Automaton for the gate in the GRC Example.}
	\label{fig:grcg}
	\end{figure}

\begin{figure}
	\centering
	\includegraphics[scale=0.5]{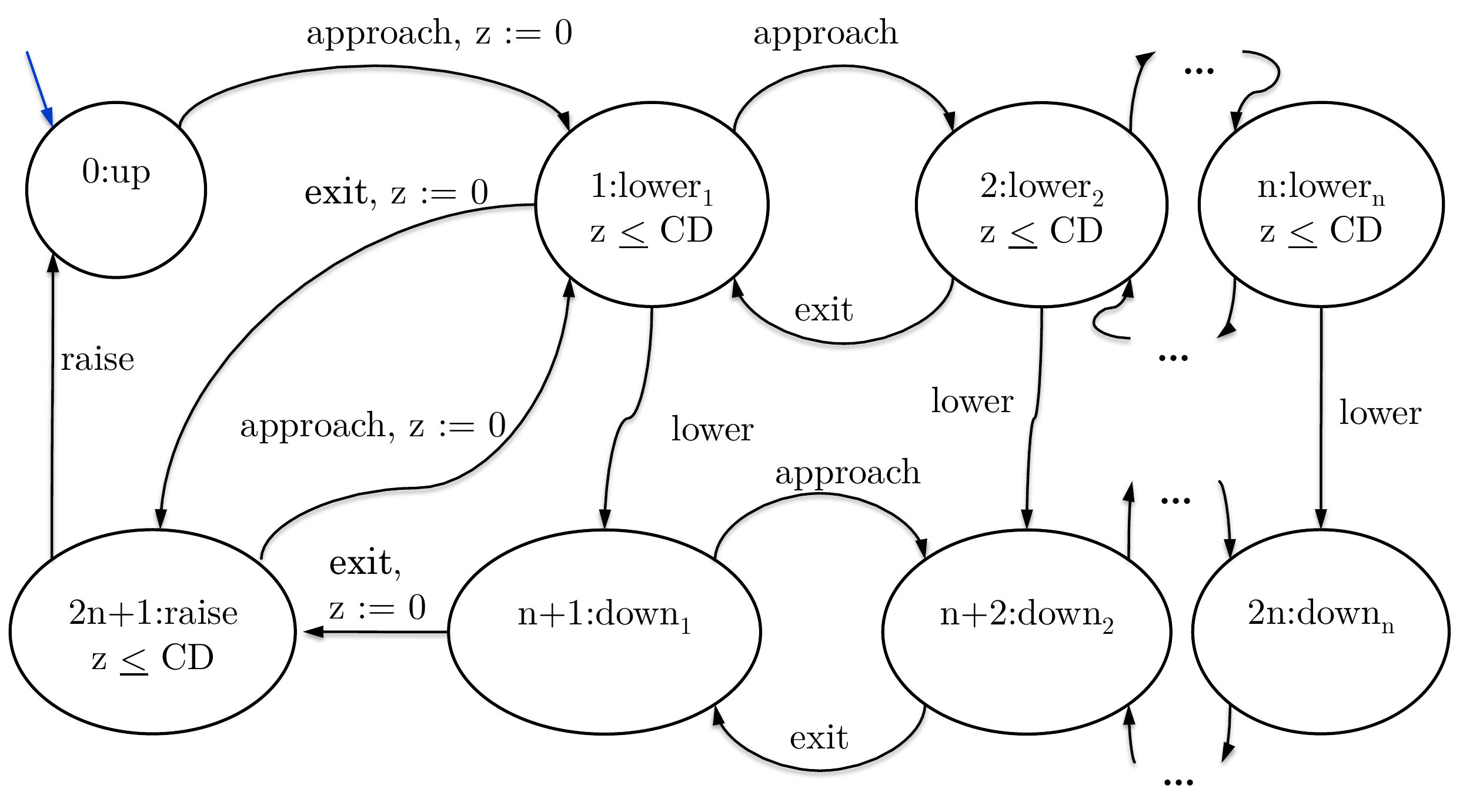}
	\caption{Timed Automaton for the controller in the GRC Example.}
	\label{fig:grcc}
	\end{figure}

	Now the figure of the gate is in Figure \ref{fig:grcg} and the controller is in Figure \ref{fig:grcc}.

	The states of the gate are:
	
	\begin{enumerate}
		\item [0.] \texttt{up:} The gate is currently up.
		\item [1.] \texttt{lower:} The gate is currently being lowered.
		\item [2.] \texttt{down:} The gate is currently down.
		\item [3.] \texttt{raise:} The gate is currently being raised.
	\end{enumerate}
	
	The states of the controller are:
	
	\begin{enumerate}
		\item State 0:\texttt{up}.  The controller has already sent the ``raise'' signal to the gate.
		\item States 1--n:$\texttt{lower}_i$.  The controller needs to send a ``lower'' signal and there are $i$ trains either approaching or in.
		\item States (n+1)--2n:$\texttt{down}_i$. The controller has sent the ``lower'' signal and there are $i$ trains either approaching or in.
		\item State (2n+1):\texttt{raise}.  The last train has left and the controller needs to send the ``raise'' signal.
	\end{enumerate}
	
	Each process (train) has one clocks $x_i$, the gate has clock $y$ and the controller has clock $z$.  Initially, each clock  is 0.  Thus, $x_i = 0$ (for all $i$) and $y=0,z=0$.  Furthermore, each process (train) is initially in state $0:far$, the gate is initially in $0:up$ and the controller is initially in $0:up$. 	
	
	\begin{remark} 
Some of these edges are unreachable, given that the \emph{safety property} holds (which it does, since we check it as our \emph{A Example}.  These edges include the $lower_i \ttrans{exit} lower_{i-1}$ when $i > 1$ for the controller, as well as the $lower \ttrans{raise} raise$ edge for the gate, since the gate is always down when train is in.  Thus, the controller will be in the down state since the gate will be down before a train exits.
	\end{remark}

\subsection{Specifications Checked}

When modifying the specifications for GRC, $p1 \ldots p_t$ are the $t$ trains, $p_{t+1}$ is the gate ($p_{g}$) and $p_{t+2}$ (notated $p_{c}$) is the controller. In TCTL formulae, Iterations over $i$ are iterations over all trains $p_i$ but do not iterate over the gate or the controller. In the Tool MES Specifications, there are $t=2$ trains. This means that in those formulae $p3$ is the gate and $p4$ is the controller 

The specifications checked on the GRC protocol are:

\noindent \textbf{AS:} It is always the case that if at least one train (process) is in the track region, the gate is always down. (Valid)

TCTL Formula: 
\begin{align} \displaystyle
\ctlAG{\bigvee_{i}(p_i = 2) \lgcif p_g = 2}
\end{align}

The $L^{rel}_{\nu,\mu}$ is included in the Tool MES Specification (with abbreviations for notational convenience).

Tool MES Specification:
\begin{lstlisting}
PREDICATE: {X}
START: X
EQUATIONS: {
1: nu X = ((p1 != 2) || (p3 == 2)) && \forall time(\AllAct(X))
}
\end{lstlisting} \parspc

\noindent \textbf{BS:}  It is always the case that if the gate is raising then the controller (when one train is approaching or in) will not want to lower the gate. (Invalid)

TCTL Formula: 
\begin{align} \displaystyle
\ctlAG{p_g = 3 \lgcif p_c \neq 1}
\end{align}

The $L^{rel}_{\nu,\mu}$ is included in the Tool MES Specification (with abbreviations for notational convenience).

Tool MES Specification:
\begin{lstlisting}
PREDICATE: {X}
START: X
EQUATIONS: {
1: nu X = ((p3 != 3) || (p4 != 1)) && \forall time(\AllAct(X))
}
\end{lstlisting} \parspc

\noindent \textbf{AL:} It is inevitable that the gate is up. (Valid)

TCTL Formula: 
\begin{align} \displaystyle
\ctlAF{p_g = 0}
\end{align}

The $L^{rel}_{\nu,\mu}$ is included in the Tool MES Specification (with abbreviations for notational convenience).

Tool MES Specification:
\begin{lstlisting}
PREDICATE: {X}
START: X
EQUATIONS: {
1: mu X = \forall time\rel[(p3==0)]((p3==0) || \AllAct(X)) && (UnableWaitInf || \exists time((p3==0)))
}
\end{lstlisting} \parspc

\noindent \textbf{BL:} It is inevitable that the train is near the gate. (Invalid)

TCTL Formula: 
\begin{align} \displaystyle
\ctlAF{\bigvee_{i}(p_i = 1)}
\end{align}

The $L^{rel}_{\nu,\mu}$ is included in the Tool MES Specification (with abbreviations for notational convenience).

Tool MES Specification:
\begin{lstlisting}
PREDICATE: {X}
START: X
EQUATIONS: {
1: mu X = \forall time\rel[((p1 == 1) || (p2 == 1))](((p1 == 1) || (p2 == 1)) || \AllAct(X)) && (UnableWaitInf || \exists time(((p1 == 1) || (p2 == 1))))
}
\end{lstlisting} \parspc

\noindent \textbf{M1:} It is always the case that if the gate is down, then it will inevitably come up (Invalid).

TCTL Formula: 
\begin{align} \displaystyle
\ctlAG{p_g = 2 \lgcif \ctlAF{p_g = 0}}
\end{align}

The $L^{rel}_{\nu,\mu}$ is included in the Tool MES Specification (with abbreviations for notational convenience).

Tool MES Specification:
\begin{lstlisting}
PREDICATE: {X, X2}
START: X
EQUATIONS: {
1: nu X = \forall time( ({p3 != 2} || X2) && \AllAct(X))
2: mu X2 = \forall time\rel[(p3 == 0)]((p3 == 0) || \AllAct(X2)) && (UnableWaitInf || \exists time((p3 == 0)))
}
\end{lstlisting} \parspc

\noindent \emph{\textbf{M2$^{*}$:} It is always the case that if the gate is down, then it will inevitably come up after 30 seconds (Invalid).}

TCTL Formula: 
\begin{align} \displaystyle
\text{No known TCTL Formula.}
\end{align}

The $L^{rel}_{\nu,\mu}$ is included in the Tool MES Specification (with abbreviations for notational convenience).

Tool MES Specification:
\begin{lstlisting}
PREDICATE: {X, X2}
START: X
EQUATIONS: {
1: nu X = \forall time( ({p3 != 2} || X2[z]) && \AllAct(X))
2: mu X2 = \forall time\rel[(p3 == 0 && z <= CWAIT)]((p3 == 0 && z <= CWAIT) || \AllAct(X2)) && (UnableWaitInf || \exists time((p3 == 0 && z <= CWAIT)))
}
\end{lstlisting} \parspc

\noindent \textbf{M3:} It is always the case that at most one train is in the region at one time (Invalid).

TCTL Formula: 
\begin{align} \displaystyle
\ctlAG{\bigwedge_{i \neq j}(p_i \neq 2 \lgcor p_j \neq 2)}
\end{align}

The $L^{rel}_{\nu,\mu}$ is included in the Tool MES Specification (with abbreviations for notational convenience).

Tool MES Specification:
\begin{lstlisting}
PREDICATE: {X}
START: X
EQUATIONS: {
1: nu X = \forall time( ({p1 != 2} || p2 != 2) && \AllAct(X))
}
\end{lstlisting} \parspc

\noindent \emph{\textbf{M4$^{*}$:} For all paths with an infinite number of actions, the gate is up until a train approaches (Valid).}

TCTL Formula: 
\begin{align} \displaystyle
\text{No known TCTL formula.}
\end{align}

The $L^{rel}_{\nu,\mu}$ is included in the Tool MES Specification (with abbreviations for notational convenience).

Tool MES Specification:
\begin{lstlisting}
PREDICATE: {X}
START: X
EQUATIONS: {
1: mu X = ((p1 == 1) || (p2 == 1)) || ((p3 == 0) && \forall time(\AllAct(X)))
}
\end{lstlisting} \parspc

\noindent \emph{\textbf{M4ap$^{*}$:} For all paths, the gate is up until a train approaches (Invalid).} 

TCTL Formula: 
\begin{align} \displaystyle
\text{No known TCTL formula.}
\end{align}

The $L^{rel}_{\nu,\mu}$ is included in the Tool MES Specification (with abbreviations for notational convenience).

Tool MES Specification:
\begin{lstlisting}
PREDICATE: {X}
START: X
EQUATIONS: {
1: mu X = \forall time\rel[((p1 == 1) || (p2 == 1))]( ((p3 == 0) || ((p1 == 1) || (p2 == 1))) && (((p1 == 1) || (p2 == 1)) || \AllAct(X))) && ( UnableWaitInf || \exists time\rel[(p3 == 0)](((p1 == 1) || (p2 == 1))))}
\end{lstlisting} \parspc

\subsection{Model in PES}

Below is the GRC Model for 2 processes specified in PES Code. The equation is the formula, and for this model the formula is the ``as'' formula.

\begin{lstlisting}
#define CCD 1
#define CGLT 2
#define CGRT 2
#define CTP 4
#define CTDL 1
#define CTDU 15
CLOCKS: {x1,x2,x3,x4}
// p1-p2 are the trains.
// p3 is the gate.
// p4 is the controller.
CONTROL: {p1,p2,p3,p4}
INITIALLY: x1 == 0 && x2 == 0 && x3 == 0 && x4 == 0
PREDICATE: {X}
START: X
EQUATIONS: {
1: nu X = ((p1 != 2) || (p3 == 2)) && \forall time(\AllAct(X))
}
INVARIANT:
	p1 == 1 -> x1 <= CTP
	p1 == 2 -> x1 <= CTDU
	p2 == 1 -> x2 <= CTP
	p2 == 2 -> x2 <= CTDU
	p3 == 1 -> x3 <= CGLT
	p3 == 3 -> x3 <= CGRT
	p4 == 1 -> x4 <= CCD
	p4 == 2 -> x4 <= CCD
	p4 == 5 -> x4 <= CCD
TRANSITIONS:
	(p1 == 0 && p4 == 0)->(p1=1, p4=1){x1,x4};
	(p1 == 0 && p4 == 5)->(p1=1, p4=1){x1,x4};
	(p1 == 0 && p4 == 1)->(p1=1, p4=2){x1};
	(p1 == 0 && p4 == 3)->(p1=1, p4=4){x1};
	(p2 == 0 && p4 == 0)->(p2=1, p4=1){x2,x4};
	(p2 == 0 && p4 == 5)->(p2=1, p4=1){x2,x4};
	(p2 == 0 && p4 == 1)->(p2=1, p4=2){x2};
	(p2 == 0 && p4 == 3)->(p2=1, p4=4){x2};
	(p1 == 1, x1 == CTP)->(p1=2){x1};
	(p2 == 1, x2 == CTP)->(p2=2){x2};
	(p1 == 2 && p4 == 1, x1 >= CTDL)->(p1=0, p4=5){x4};
	(p1 == 2 && p4 == 3, x1 >= CTDL)->(p1=0, p4=5){x4};
	(p1 == 2 && p4 == 2, x1 >= CTDL)->(p1=0, p4=1);
	(p1 == 2 && p4 == 4, x1 >= CTDL)->(p1=0, p4=3);
	(p2 == 2 && p4 == 1, x2 >= CTDL)->(p2=0, p4=5){x4};
	(p2 == 2 && p4 == 3, x2 >= CTDL)->(p2=0, p4=5){x4};
	(p2 == 2 && p4 == 2, x2 >= CTDL)->(p2=0, p4=1);
	(p2 == 2 && p4 == 4, x2 >= CTDL)->(p2=0, p4=3);
	(p3 == 2 && p4 == 5)->(p3=3, p4=0){x3};
	(p3 == 1 && p4 == 5)->(p3=3, p4=0){x3};
	(p3 == 0 && p4 == 1)->(p3=1, p4=3){x3};
	(p3 == 3 && p4 == 1)->(p3=1, p4=3){x3};
	(p3 == 0 && p4 == 2)->(p3=1, p4=4){x3};
	(p3 == 3 && p4 == 2)->(p3=1, p4=4){x3};
	(p3 == 3, x3 == CGRT)->(p3=0);
	(p3 == 1, x3 == CGLT)->(p3=2);
\end{lstlisting}

\section{LEADER Election}
\label{s:leader}

	Additional information about the protocol was obtained by \cite{romijn-a-timed-2001,verdejo-the-leader-2000}.
	
\subsection{Overview}

	The LEADER Protocol	involves $n$ processes that are electing a leader amongst themselves.  They do this by asking a fellow process to be their ``parent".  The process that is being asked to be their parent either accepts the request or denies it.   After the request, the request-responding pair is resolved my making the smaller id process the parent of the other.  Each process can have at most one parent but can have any number of child processes.  Each request is executed instantaneously.  When the election protocol is ``finished'', the ``root" process is the leader.	  
	
	Parameters:
	
\begin{itemize}
	\item $\mathbf{CPD}$ (PERIOD).  This is the length of time a request-respond cycle can take for a process requesting a parent and then being accepted or denied.  The default value is 2.
\end{itemize}

\subsection{Timed Automata Modeling}
	
	These are the actions which the processes are ``synchronized'' on. 	
	\begin{itemize}
		\item $req_{ij}$, here process $i$ asks $j$ to be its parent.  Note that the request always resolves where $i$ becomes $j$'s parent if $i < j$ and if $i > j$, $j$ becomes i's parent.  We do not allow $i = j$.
	\end{itemize}
	
\begin{figure}
	\centering
	\includegraphics[scale=0.5]{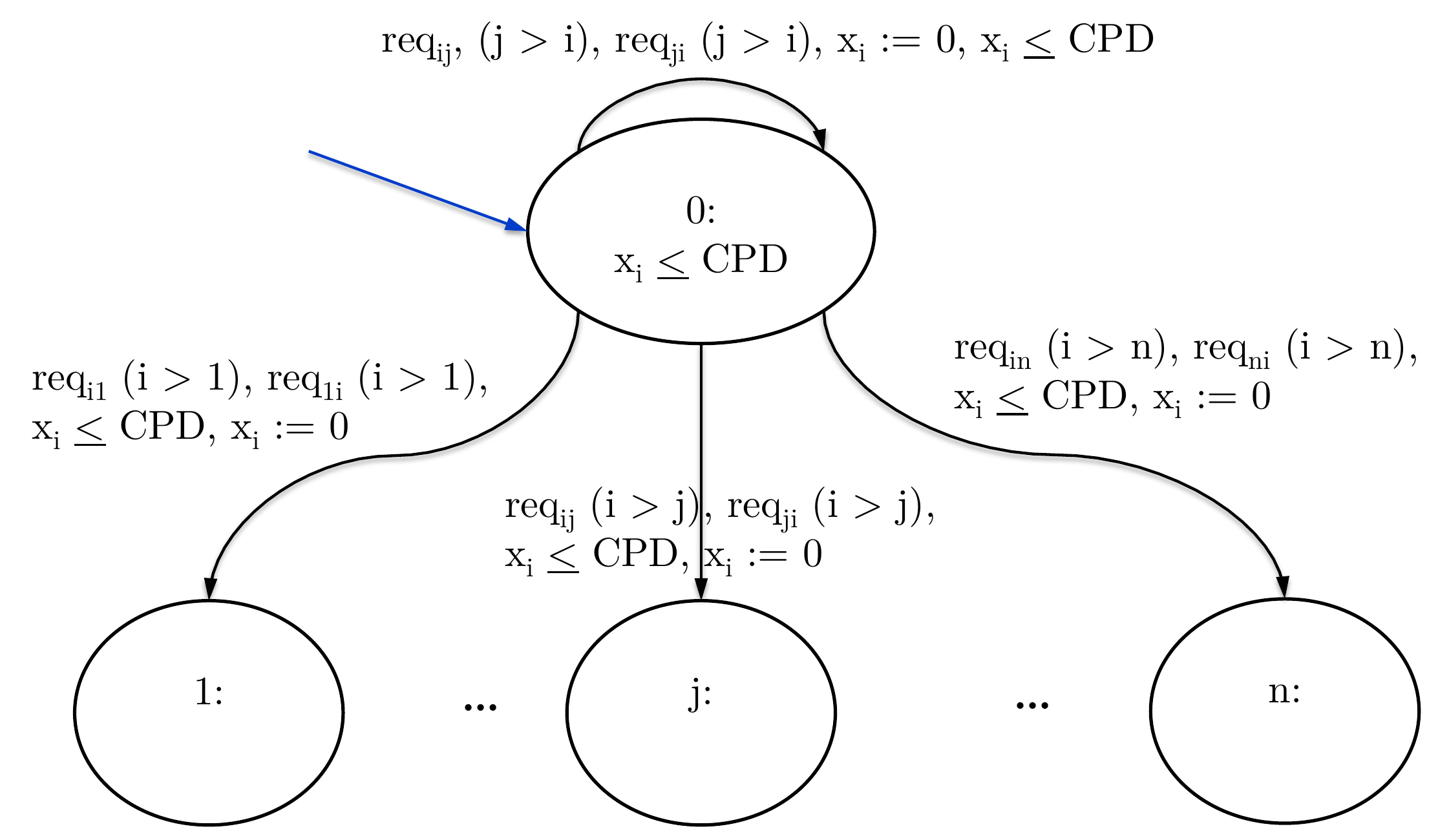}
	\caption{Timed Automaton for a Process in the LEADER Example.}
	\label{fig:leaderp}
	\end{figure}

	The Timed Automaton Figure for a single process is in Figure \ref{fig:leaderp}. 
		
	The states of Process $i$ are as follows:
	
	\begin{enumerate}		
		\item [0.] \texttt{0:} The process has no parent.
		\item [i.] \texttt{i:} The process has process $i$ as its parent.  Here $i$ ranges from $1\text{--}n$. 
	\end{enumerate}

	Given the instantaneous resolution of requests, there is no need to store a global variable.  The local variable of the process's parent is stored with the state.

\begin{remark}
The LEADER election process model here follows the model in \cite{wang-efficient-verification-2003}.  This model simplifies things from the models in \cite{romijn-a-timed-2001,verdejo-the-leader-2000}:

\begin{itemize}
	\item It assumes the network topology is the complete graph of processes.
	\item Then the labeling of numbers is a ``symmetry'': although any process can request a parent, requests are always resolved where the smaller process id becomes the parent of the other, regardless of who made the initial request.
	\item The non-determinism is implemented in a manner that there are no cyclic conflicts.  All the resolved outcomes are still available.
\end{itemize}

\end{remark}	

	Each process has one clock $x_i$.  Initially, each clock is 0 ($x_i = 0$ for all $i$) and each process is initially in state 0 (has no parent).
	
	\begin{remark}[Timelock] This model has a timelock (with the state when a leader is elected). To ensure that the timed automaton is timelock-free, an additional final state where time can diverge is added. This state can be transitioned to when a leader is elected. \end{remark}

\subsection{Specifications Checked}

In the specifications, the processes are $p_1 \ldots p_l$ for the l processes. In addition, there is a variable $p$ that is initially 0 but becomes $1$ when the leader election process is finished.

The specifications checked on the LEADER protocol are:

\noindent \textbf{AS:} At any time, each process either has no parent or has a parent with a smaller process id (and thus the first process has no parent at all times). (Valid)

TCTL Formula: 
\begin{align} \displaystyle
\ctlAG{\bigwedge_{i}(p_i < i)}
\end{align}

The $L^{rel}_{\nu,\mu}$ is included in the Tool MES Specification (with abbreviations for notational convenience).

Tool MES Specification:
\begin{lstlisting}
PREDICATE: {X}
START: X
EQUATIONS: {
1: nu X = ((p1 < 1)
	&&(p2 < 2)
) && \forall time(\AllAct(X))
}
\end{lstlisting} \parspc

\noindent \textbf{BS:} At any moment, at least three processes do not have parents.	(Invalid)

TCTL Formula: 
\begin{align} \displaystyle
\ctlAG{\bigwedge_{i \neq j \neq k}(p_i = 0 \lgcand p_j = 0 \lgcand p_k = 0)}
\end{align}

The $L^{rel}_{\nu,\mu}$ is included in the Tool MES Specification (with abbreviations for notational convenience).

Tool MES Specification:
\begin{lstlisting}
START: X
EQUATIONS: {
1: nu X = ((p1 == 0 && p2 == 0 && p3==0)
) && \forall time(\AllAct(X))
}
\end{lstlisting} \parspc

\noindent \textbf{AL:}  It is inevitable that the first process is elected the leader. (Valid)

TCTL Formula: 
\begin{align} \displaystyle
\ctlAF{p_1 = 0 \lgcand \bigwedge_{i \neq 1}(p_i \neq 0)}
\end{align}

The $L^{rel}_{\nu,\mu}$ is included in the Tool MES Specification (with abbreviations for notational convenience).

Tool MES Specification:
\begin{lstlisting}
PREDICATE: {X}
START: X
EQUATIONS: {
1: mu X = \forall time\rel[((p1 == 0) && (p2 != 0))](((p1 == 0) && (p2 != 0)) || \AllAct(X)) && (UnableWaitInf || \exists time(((p1 == 0) && (p2 != 0))))
}
\end{lstlisting} \parspc

\noindent \textbf{BL:} It is inevitable that the third processes' parent is the second process. (Invalid)

TCTL Formula: 
\begin{align} \displaystyle
\ctlAF{p_3 = 1}
\end{align}

The $L^{rel}_{\nu,\mu}$ is included in the Tool MES Specification (with abbreviations for notational convenience).

Tool MES Specification:
\begin{lstlisting}
PREDICATE: {X}
START: X
EQUATIONS: {
1: mu X = \forall time\rel[( p3 == 1)](( p3 == 1) || \AllAct(X)) && (UnableWaitInf || \exists time(( p3 == 1)))
}
\end{lstlisting} \parspc

\noindent \emph{\textbf{M1$^{*}$:} For all paths, a the second process cannot have a child until it has a parent. (Invalid)}

TCTL Formula: 
\begin{align} \displaystyle
\text{No known TCTL formula.}
\end{align}

The $L^{rel}_{\nu,\mu}$ is included in the Tool MES Specification (with abbreviations for notational convenience).

Tool MES Specification:
\begin{lstlisting}
PREDICATE: {X}
START: X
EQUATIONS: {
1: mu X = \forall time\rel[(p2 != 0)]( (((p1 != 2) && (p2 != 2)) || (p2 != 0)) && ((p2 != 0) || \AllAct(X))) && ( UnableWaitInf || \exists time\rel[((p1 != 2) && (p2 != 2))]((p2 != 0)))}
\end{lstlisting} \parspc

\noindent \textbf{M2:} It is always the case that if the third process is assigned a parent (chosen to not be leader), then it will not be the leader. (Valid)

TCTL Formula: 
\begin{align} \displaystyle
\ctlAG{p_3 \neq 0 \lgcif \ctlAG{p_3 \neq 0}}
\end{align}

The $L^{rel}_{\nu,\mu}$ is included in the Tool MES Specification (with abbreviations for notational convenience).

Tool MES Specification:
\begin{lstlisting}
PREDICATE: {X,X2}
START: X
EQUATIONS: {
1: nu X = \forall time((({p3 == 0} || X2)) && \AllAct(X))2: nu X2 = \forall time(((p3 != 0)) && \AllAct(X2))}
\end{lstlisting} \parspc

\noindent \emph{\textbf{M3$^{*}$:} It is possible that it takes longer than 3 time units to elect a leader. (Valid)}

TCTL Formula: 
\begin{align} \displaystyle
\tctlEG{p = 0}{> 3}
\end{align}

The $L^{rel}_{\nu,\mu}$ is included in the Tool MES Specification (with abbreviations for notational convenience).

Tool MES Specification:
\begin{lstlisting}
PREDICATE: {X, X2}
START: X
EQUATIONS: {
1: mu X = X2[z]2: mu X2 = \exists time(((p == 0 && z>=3)) || \ExistAct(X2))}
INVARIANT:
	p1 == 0 && p==0 -> x1 <= CPD
	p2 == 0 && p==0 -> x2 <= CPD
TRANSITIONS:
	(p2 == 0 && p1 == 0, x2 <= CPD && x1 <= CPD)->(p2 = 1){x2, x1};
	(p==0 && p1==0 && p2!=0)->(p=1){x1, x2};
\end{lstlisting} \parspc

\noindent \emph{\textbf{M4$^{*}$:} For all paths, in at most three votes, a leader is elected.  (Valid for four or fewer processes, invalid for five or more processes.)}  

TCTL Formula: 
\begin{align} \displaystyle
\text{No known TCTL formula.}
\end{align}

The $L^{rel}_{\nu,\mu}$ is included in the Tool MES Specification (with abbreviations for notational convenience).

Tool MES Specification:
\begin{lstlisting}
PREDICATE: {X}
START: X
EQUATIONS: {
1: nu X = \forall time( (p == 1) || \AllAct(\forall time( (p == 1) || \AllAct(\forall time( (p == 1) || \AllAct(\forall time( (p == 1) || \AllAct(\forall time( (p == 1))) )) )) )) )
}
\end{lstlisting} \parspc

	\begin{remark} The more typical \textbf{A Example} specification is that at all times, at least one process has no parent.  This is a corollary of the \textbf{A Example} specification that we check.  While this is a reasonable specification to check, we check a stronger one since this specification is not invalidated by common bugs on the more simpler LEADER model that we took.  Of course, errors can always be introduced to invalidate this specification, but this weaker specification is
	\end{remark}
	
With these specifications, the process can by symmetric, but in this specification, it is assumed that the lower numbered process becomes the leader. This results in shorter specifications.
	
\subsection{Model in PES}

Below is the LEADER Model for 2 processes specified in PES Code. The equation is the formula, and for this model the formula is the ``as'' formula.

\begin{lstlisting}
#define CPD 2
CLOCKS: {x1,x2}
CONTROL: {p1,p2, p}
PREDICATE: {X}
START: X
EQUATIONS: {
1: nu X = ((p1 < 1)
	&&(p2 < 2)
) && \forall time(\AllAct(X))
}
INVARIANT:
	p1 == 0 && p==0 -> x1 <= CPD
	p2 == 0 && p==0 -> x2 <= CPD
TRANSITIONS:
	(p2 == 0 && p1 == 0, x2 <= CPD && x1 <= CPD)->(p2 = 1){x2, x1};
	(p==0 && p1==0 && p2!=0)->(p=1){x1, x2};
\end{lstlisting}

\newpage
\bibliographystyle{plainnat}
\addcontentsline{toc}{section}{References}
\bibliography{TABenchmarkReport}

\end{document}